\documentclass[12pt,a4paper]{article}
\pdfoutput=1 % 
\usepackage[utf8]{inputenc}
\usepackage[T1]{fontenc}
\usepackage{fullpage}
\usepackage[sort&compress,numbers]{natbib}
\usepackage{subcaption}
\usepackage[english]{babel}
\usepackage{amsmath}
\usepackage{amsthm}
\usepackage{amsfonts}
\usepackage{amssymb}
\usepackage{bbold}
\usepackage{mathrsfs}
\usepackage{mathbbol}
\usepackage[toc,page]{appendix}
\usepackage{subcaption}
\usepackage{graphicx}
\usepackage{bbding}
\usepackage{esvect}
\usepackage{commath}
\usepackage{enumitem}
\usepackage[affil-it]{authblk}
\usepackage{siunitx}
\usepackage{braket}
\usepackage{phaistos}
\usepackage{multicol}
\usepackage{overpic}
\usepackage{comment}
\usepackage{float}
\usepackage{overpic}

\usepackage{xcolor}

\usepackage{jheppub}
\usepackage{tikz}

\usepackage{algorithm}
\usepackage[noend]{algpseudocode}

\usepackage{environ}
\NewEnviron{myequation}{%
\begin{equation*}
\scalebox{1.7}{$\BODY$}
\end{equation*}
}

\usepackage{bbm}% Added by cindy for that nice \mathbbm{1}

\def\ket#1{{|{#1}\rangle}} 

\title{A Compact Framework for Analyzing Asynchronous Entanglement Distribution in Quantum Networks}

\author[a]{Emma Hughes,}
\author[a]{William Munizzi,}
\author[a,b]{Prineha Narang}

\affiliation[a]{Division of Physical Sciences, College of Letters and Science, University of California, Los Angeles, CA 90095 USA}
\affiliation[b]{Department of of Electrical and Computer Engineering, University of California, Los Angeles, CA 90095 USA}

\emailAdd{emmahughes@g.ucla.edu}
\emailAdd{wmunizzi17@g.ucla.edu}
\emailAdd{prineha@ucla.edu}

\abstract{This work introduces a compact framework for analyzing asynchronous entanglement distribution protocols under realistic error models. We focus on two contemporary protocols: sequential, where entanglement is established one node at a time, and parallel, where all nodes attempt to generate entanglement simultaneously. We derive an analytical expression for the fidelity of distributed entangled states, showing that the fidelity depends only on the total time all qubits spend in memory, rather than the individual memory times for each qubit. This result distills the complex dynamics of entanglement distribution into a compact accessible form, providing an scalable tool for evaluating protocol efficiency. Using this lightweight framework, we analyze the performance of parallel and sequential protocols, demonstrating that parallel distribution consistently outperforms sequential and highlighting the potential of parallel protocols for practical quantum network implementations.}

\begin{document} 
\maketitle
\flushbottom

%%%%%%%%%%%%%%%%%%%%%%%%%%%%%%%%%%%%%%%%%%%%%%%%%%%%%%%%%%%%%%%%%%%%%%%%%%%%%%%%%%%%%%%

\section{Introduction}
Distributing entangled states is fundamental to many proposed applications for quantum networks~\cite{vanmeter2014quantum, castelvecchi2018quantum, kimble2008quantum, wehner2018quantum}, such as quantum computing~\cite{broadbent2016quantum, beals2013efficient, ge2018distributed, fitzsimons2017private}, quantum key distribution~\cite{BB84}, and an eventual quantum internet~\cite{Kimble_2008,Wehner_2018}. However, given the extreme susceptibility of contemporary quantum devices to noise, decoherence and information loss threaten the near-term realization of these technological ambitions. Photonic quantum network schemes, in particular, suffer substantial information loss occurring throughout the optical transmission of states. In an attempt to remedy this loss, numerous entanglement distribution protocols have been proposed~\cite{Briegel_1998,Munro_2015,Azuma_2022,Haldar_2024,10637640,PhysRevA.77.022308,10679782}. In this work, we evaluate leading entanglement distribution protocols under realistic noise conditions using two key metrics: the fidelity of the final distributed entangled states and the rate at which high fidelity (near perfect) entangled states can be distilled through entanglement purification~\cite{Bennett1996purification,Deutsch1996privacy, purification,Rozpedek2018distillation,Zhao2021LOCCNet,Jansen2022clifford, PhysRevResearch.5.033171}, known as the hashing rate~\cite{hashing-rate}, derived by Bennett et al. 

To distribute a bipartite entangled state between a sender and a receiver separated by a large distance, quantum repeaters can be introduced along the communication channel. Rather than transmitting entangled photons directly over the entire distance (which would suffer significant losses) entanglement is first established over shorter segments between neighboring nodes, forming \textit{entanglement links} between the nodes. These nodes, consisting of the sender, receiver, and intermediate quantum repeaters, form a linear, entanglement-based quantum network. Through entanglement swapping at the repeater nodes, these short-distance entanglement links can be used to generate high-fidelity long-distance entanglement between the sender and receiver.

In this work, each entanglement link between two nodes is generated by producing an entangled photon pair at one node and transmitting one photon to the other. This is different from common schemes in which entangled photons are generated midway between the repeaters, which, despite being more efficient, has an additional hardware overhead which is impractical for scaling to larger networks~\cite{cisco-analysis-asynchronous, Pouryousef_2023}.

Once a given repeater has established entanglement with both of its neighbors, it performs an entanglement swapping operation which establishes a single entangled pair between the distant neighbors. Entanglement swapping takes place on all the repeaters until the sender and receiver are connected by a single long-distance entanglement link.

It is commonly assumed when analyzing entanglement distribution protocols that all nodes have global-knowledge and entanglement generation attempts can be made with perfect synchronization throughout the network. These protocols are referred to as synchronous or time-slotted protocols. While this simplifies design, it demands a central controller and perfect synchronization—an increasingly unrealistic requirement for larger networks~\cite{cisco-analysis-asynchronous, Pouryousef_2023}.

In contrast, this work focuses on asynchronous entanglement distribution protocols~\cite{pouryousef2024analysisasynchronousprotocolsentanglement,9963998,yang2024asynchronous}, where each node independently attempts to generate entanglement with its neighbor. This removes the need for centralized control and allows for the possibility of distributed routing strategies.

\section{Entanglement Distribution and Noise in Quantum Networks}\label{NoiseModel}

In this work, we simulate and analyze asynchronous routing protocols used to distribute a Bell pair across distant nodes in a linear quantum network. We operate under the assumption that each node in the network has the ability to generate entanglement, as well as to perform entanglement swapping operations. The nodes themselves are considered to be evenly spaced, and physically connected using optical fibers. In this section we review the necessary details behind Bell state distribution, subject to the above constraints. Furthermore, we describe the greatest factors contributing to signal loss and unsuccessful state distribution in asynchronous routing protocols.

The process of Bell state distribution begins by selecting two end nodes, one of which serves as the sender node (which initiates the communication), and the other as the receiver node (which receives the signal). Between these two end nodes exists a collection of intermediate repeater nodes, each of which serves to mitigate optical loss occurring throughout the protocol. We first consider the $2$-qubit Bell state $\ket{\phi^+}$, defined
\begin{equation}\label{bellState}
    \ket{\phi^+} = \frac{1}{\sqrt{2}}( \ket{00}+\ket{11}),
\end{equation}
that we wish to distribute between the end nodes in our linear network. One of the qubits from the pair is transmitted, from sender node to receiver node, via a sequence of entanglement swapping operations applied at each repeater node along the line connecting the end nodes in the linear quantum network.

Distributing entanglement across distant nodes in a quantum network relies on two key operations: entanglement generation and entanglement swapping. Entanglement generation refers to the distribution of an entangled pair between adjacent nodes, thereby establishing entanglement between the nodes. For adjacent nodes $A$ and $B$ in the quantum network, separated by a distance $L$, the entanglement generation protocol begins by generating a pair of entangled photons in node $A$. One photon is then sent to node $B$ through an optical fiber. If the attempted entanglement generation is successful, classical communication is sent back from $B$ to $A$ to confirm receipt of the entangled photon. At the end of the event, the photon in node $A$ spends time $t= 2L/c$ in memory, and the qubit at node $B$ spends time $t= L/c$ in memory.

Entanglement swapping involves two pairs of entangled photons $(A, B)$ and $(C, D)$. A Bell state measurement is performed on photons $(B, C)$, resulting in the two shorter-distance entanglements to be effectively swapped into one longer distance entanglement between photons $(A, D)$. Once entanglement is successfully established on both sides of the node, i.e. the node has both successfully received an entangled photon and sent an entangled photon to its neighboring node, an entanglement swapping operation is performed, establishing entanglement between its neighboring nodes.

In this work we consider two asynchronous distribution protocols, the sequential protocol and the parallel protocol. This transmission process is inherently noisy, arising from various sources throughout the entanglement distribution process, which we now review and analyze.

\subsection{An Error Model for Entanglement-Based Quantum Networks}

Contemporary models for quantum networks rely on optical fibers to transmit information between distant nodes. As a result, a central focus is on understanding and characterizing the specific noise and loss associated with optical hardware. One source of noise arises from an exponential signal attenuation of with increasing propagation distance. We characterize this effect by the channel efficiency~\cite{requsim}, defined
\begin{equation}
    \eta_{ch}(L) \equiv e^{\frac{-L}{L_{att}}}, 
\end{equation}
where $L$ denotes the length the photon must travel between nodes, and $L_{att}$ is the attenuation length. For optical fibers operating at Telecom wavelengths, the primary focus of this work, the value of $L_{att}$ is typically around $22$ km.

Another significant source of loss in photonic quantum networks arises from the need to store quantum information, carried by the photons, for ready access. The storage process, known as quantum memory, is typically realized through some light-matter interaction, e.g. encoding the photon into an atomic ensemble. While quantum memories enable strong utility for information processing in a quantum network, they are subject to time-dependent dephasing that degrades the stored information. For a quantum network described by the density matrix $\rho$, the dephasing of the $i^{th}$ qubit, stored in memory for time $t$, is given by
\begin{equation}
    \mathcal{E}_Z^{(i)}(t)\rho = (1 - \lambda(t))\rho + \lambda(t) Z^{(i)} \rho Z^{(i)},
\end{equation}
where $Z^{(i)}$ indicates Pauli $Z$ action on the $i^{th}$ qubit, and $\lambda(t)$ is a function of time given by
\begin{equation}
\lambda(t) = \frac{1 - e^{-t/T_{\text{dp}}}}{2},
\end{equation}
which depends on the memory dephasing time $T_{\text{dp}}$ of the particular hardware platform. 

Successful entanglement swapping depends on the ability to perform high-fidelity Bell state measurement (BSM) on transmitted quantum states. A Bell state measurement consists of a $2$-qubit measurement, followed by the projection of their combined state onto one of four maximally entangled Bell states, specifically
\begin{equation}\label{BellStates}
\begin{split}
    \ket{\phi^{\pm}} &\equiv \frac{1}{2} \left(\ket{00} \pm \ket{11} \right),\\
    \ket{\psi^{\pm}} &\equiv \frac{1}{2} \left(\ket{01} \pm \ket{10} \right).\\
\end{split}
\end{equation}
When performing entanglement swapping on noisy hardware, perfectly distinguishing each Bell state in Eq.\ \eqref{BellStates} and performing the necessary projective measurement is rarely achieved without error. Imperfect Bell state measurement occurs due to a variety of factors, including photon detector limitations, decoherence, and the indistinguishability of states in certain optical realizations. In this work, we model imperfect Bell state measurement by applying a $2$-qubit depolarizing channel $\mathcal{E}_w^{(i,j)}$ to the qubits $i$ and $j$ being measured, followed by an ideal Bell state measurement. We define the depolarizing channel $\mathcal{E}_w^{(i,j)}$ as
\begin{equation}\label{eqn:bsm-error}
\mathcal{E}_w^{(i,j)}\left(\lambda_{\text{BSM}},\rho \right) = \lambda_{\text{BSM}} \rho + \frac{1 - \lambda_{\text{BSM}}}{4} \left(\text{tr}_{i,j} \rho \right) \otimes \mathbb{1}^{(i,j)},
\end{equation}
where $0 \leq \lambda_{BSM} \leq 1$ is the BSM ideality parameter, with $\lambda_{BSM} = 1$ corresponding to perfect measurement. The object $\text{tr}_{i,j} \rho$ in Eq.\ \eqref{eqn:bsm-error} denotes a partial trace on $\rho$ after tracing out (discarding) the information stored in the measured qubits $i$ and $j$.

The primary sources of error impacting entanglement distribution in photonic quantum networks are optical fiber loss, dephasing in quantum memory, and imperfect Bell state measurements. We provided analytic expressions to capture the effects of these noise sources within a simulated quantum network. These noise models offer a realistic foundation for determining the performance of asynchronous distribution protocols on near-term quantum hardware. In the next section, we utilize the above noise models to derive a closed-form expression for output Bell state fidelity in the general case.

\subsection{A Closed Form Expression for Fidelity in an $n$-Qubit Network}

In this section, we derive a closed expression for the fidelity of an entangled state distributed by an $n$-qubit quantum network using the error model presented in the previous section. Throughout this work, we use $N$ to denote the number of network nodes and $n=2(N-1)$ to represent the number of qubits. We first demonstrate that dephasing error depends only on the total time spent in memory across all qubits, and grows exponentially as the ratio of this cumulative time over the memory dephasing time. We then combine the effect of imperfect Bell state measurement error to arrive at a general expression for the fidelity of a Bell state prepared using entanglement swapping in an $n$-qubit quantum network. 

To demonstrate that the dephasing error depends only on the total time spent in memory, we firstly consider a quantum network of four qubits. The network consists of two Bell pairs $(\ket{\phi^+} \bra{\phi^+})_{AB}$ and $(\ket{\phi^+} \bra{\phi^+})_{CD}$, shared between qubits $A$ and $B$, and qubits $C$ and $D$ respectively. We allow each qubit to remain in memory for some fixed amount of time. By performing a Bell state measurement on qubits $B$ and $C$, a new Bell pair, $(\ket{\phi^+} \bra{\phi^+})_{AD}$, can be generated. The effect of dephasing on the fidelity $F_{AD}$ of the distributed Bell pair is derived in Appendix~\ref{app-dephase-4}, and is given by
\begin{equation}\label{eq:FidelityExpression-4Qubits}
F_{AD} = \alpha_{AB} \, \alpha_{CD} + \beta_{AB}\, \beta_{CD},
\end{equation}
where $\alpha_{XY}$ and $\beta_{XY}$ are probabilistic parameters, generically defined for any Bell pair between qubits $X$ and $Y$. The parameters $\alpha_{XY}$ and $\beta_{XY}$ are given by
\begin{align}\label{eq:alpha-beta}
\alpha_{XY} &= 1 - p_X - p_Y + 2\,p_X \, p_Y \notag \\
\beta_{XY} &= p_X + p_Y - 2\,p_X \, p_Y,
\end{align}
where $p_k$ denotes the probability of the $k^{th}$ qubit experiencing dephasing after $t_k$ time, given by 
\begin{equation}\label{eq:px}
p_{k} = \frac{1 - e^{-t_{k}/T_{dp}}}{2}.
\end{equation}

Combining Eqs.\ \eqref{eq:FidelityExpression-4Qubits}, \eqref{eq:alpha-beta}, and \eqref{eq:px} we derive a simplified form for $F_{AD}$, specifically
\begin{equation}\label{Dephase4}
F_{AD} = \frac{1}{2}\left(1 + e^{\frac{-T}{T_{dp}}} \right),
\end{equation}
where $T_{dp}$ is again the memory dephasing time determined by the technological capabilities of the hardware, and $T$ denotes the cumulative time spent in memory by all four qubits given as the sum
\begin{equation}\label{TotalTime4}
    T=t_A+t_B+t_C+t_D.
\end{equation}
Importantly, the final fidelity of the $AD$ Bell state depends only on $T$, while the distribution of the time spent in memory among individual qubits is irrelevant.

Extending this result further, we now consider a $6$-qubit network comprised of three Bell pairs $\left(\ket{\phi^+}\bra{\phi^+} \right)_{AB}$, $\left(\ket{\phi^+}\bra{\phi^+} \right)_{CD}$, and now $\left(\ket{\phi^+}\bra{\phi^+} \right)_{EF}$ as well. We first perform an entanglement swapping event, as before, generating $\left(\ket{\phi^+}\bra{\phi^+} \right)_{AD}$. We then implement entanglement swapping between $\left(\ket{\phi^+}\bra{\phi^+} \right)_{AD}$ and $\left(\ket{\phi^+}\bra{\phi^+} \right)_{EF}$, performing a Bell state measurement on qubits $D$ and $E$, resulting in the state $\left(\ket{\phi^+}\bra{\phi^+} \right)_{AF}$. The fidelity $F_{AF}$ of the distributed Bell state, $\left(\ket{\phi^+}\bra{\phi^+} \right)_{AF}$, is derived in Appendix~\ref{app-dephase-n}, and has the form
\begin{equation}\label{eq:FidelityExpression-6Qubits}
F_{AF} = (\alpha_{AB} \, \alpha_{CD} + \beta_{AB} \, \beta_{CD}) \,  (\alpha_{EF} + \beta_{EF}).
\end{equation}

Combining Eqs.\ \eqref{eq:alpha-beta} and \eqref{eq:px} with  \eqref{eq:FidelityExpression-6Qubits}, we can simplify $F_{AF}$ to the following
\begin{equation}\label{Dephase6}
F_{AF} = \frac{1}{2}\left(1 + e^{\frac{-T}{T_{dp}}} \right),
\end{equation}
where, as in Eq.\ \eqref{TotalTime4}, $T$ denotes the total time spent in memory by all qubits
\begin{equation}
    T = t_A+t_B+t_C+t_D+t_E+t_F.
\end{equation}
As was the case in Eq.\ \eqref{Dephase4}, we again observe that the final Bell state fidelity depends only on the cumulative time spent in memory by all qubits. 

Since, for a repeater quantum network,  expanding to a higher number of nodes involves iteratively appending Bell pairs and performing subsequent entanglement swapping operations, the fidelity expressions derived in Eqs.\ \eqref{Dephase4} and \eqref{Dephase6} apply generically for all $n$. Otherwise stated, given a repeater quantum network of $n$ qubits $\{a_1, ..., a_n\}$, the Fidelity of a Bell state $\left(\ket{\phi^+}\bra{\phi^+} \right)_{a_1,a_n}$ generated between $a_1$ and $a_n$ via entanglement swapping events is
\begin{equation}\label{eq:fid-dephase-gen}
F_{a_1,a_n} =  \frac{1}{2}\left(1+e^{\frac{-T}{T_{dp}}} \right).
\end{equation}
The total time spent in memory $T$ among all qubits is the sum 
\begin{equation}\label{eq:time-sum}
    T = \sum_iT_{a_i}.
\end{equation}
Eq.\ \eqref{eq:fid-dephase-gen} demonstrates that, regardless of network size, the final Bell state fidelity depends only on the total quantum memory time, and is independent of how that time is distributed across individual qubits in the network.

A final element we must consider is the depolarizing effect of Bell state measurement errors in a network of arbitrary size. Each time an entanglement swapping operation is performed, with efficiency $\lambda_{BSM}$, the fidelity of the generated $\phi^+$ state acquires an ideality factor of $\lambda_{BSM}$ (see Appendix~\ref{app-bsm-error} for full derivation). Moreover, dephasing error and depolarization error are independent, and therefore the magnitude of one has no impact on the magnitude of the other. Therefore a Bell state generated across an $n$-qubit network, using $m$ entanglement swapping operations, will have fidelity
\begin{equation}\label{eq:fid-general}
    F_{a_1,a_n} = \frac{(\lambda_{BSM})^m}{2}(1+e^{\frac{-T}{T_{dp}}}),
\end{equation}
where $m$ denotes the number of Bell state measurements needed to perform all entanglement swaps. In linear networks $m=N-2$, i.e. entanglement swapping must be performed on all quantum repeaters in the $N$-node network. 

Eq.\ \eqref{eq:fid-general} provides a general expression for Bell state fidelity, prepared using $m$ entanglement swapping events in an $n$-qubit repeater network. The dependence of this fidelity expression on the total memory time $T$, as opposed to individual qubit memory times, significantly simplifies the computation. Instead of relying on memory timing histories to compute fidelity, as in previous work~\cite{cisco-analysis-asynchronous}, Eq.\ \eqref{eq:fid-general} provides a compact analytical formulation that enables efficient analysis at substantially larger qubit scales, as we demonstrate in Section \ref{ComparisonSection}.

In this section, we derived a general closed expression for the fidelity of a Bell state prepared via entanglement swapping in an $n$-qubit quantum network which incorporates the errors of fiber loss, memory dephasing and imperfect Bell state measurements. Firstly, we demonstrated that the dephasing error depends only on the total time all qubits spend in memory, rather than on the distribution of individual quantum memory times. This was demonstrated through an expression for the fidelity of the final distributed Bell state with the only variables being the total time in memory, $T$, and the memory dephasing time $T_{dp}$. The total time spent in memory during the protocol is affected by fiber loss and can be obtained through simple classical simulation. Our final fidelity expression, Eq.\ \eqref{eq:fid-general}, accounts for depolarization from imperfect Bell state measurements as well as dephasing error from quantum memories.

\subsection{Performance Evaluation Metrics}\label{sec:metrics}

We now offer a set of performance metrics for evaluating the success of entanglement distribution used for our experiments in Section~\ref{sec:effic-analysis}. First, when assessing the success of generating a desired final Bell state $\ket{\phi^+}$, we compute the fidelity of $\ket{\phi^+}$ with the experiment output state $\rho$. Since $\ket{\phi^+}$ is pure, we can compute this fidelity as
\begin{equation}\label{Fidelity}
    F (\rho, \, \phi^+)= \bra{\phi^+} \, \rho \, \ket{\phi^{+}}. 
\end{equation}
Eq.\ \eqref{Fidelity} effectively computes how closely the output $\rho$ resembles $\ket{\phi^+}$, with values $0 \leq F (\rho, \, \phi^+) \leq 1$.

Another important metric for evaluating the success of entanglement distribution is the hashing rate $R_H$, which quantifies the rate at which high-fidelity Bell pairs can be distilled from the noisy output states of a quantum network. The process of entanglement distillation considers $p$ imperfect Bell pairs of fidelity $F$, and attempts to extract from them $q$ purified (near perfect) Bell pairs, where $q \leq p$. The yield $Y$ as defined by Bennett et al.~\cite{Bennett1996purification}, characterizes the number of near perfect Bell pairs distilled per input Bell pair of fidelity $F$, and is computed using
\begin{equation}
Y_H = \frac{q}{p} = 1 + F \log_2(F) + (1 - F) \log_2\left(\frac{1 - F}{3}\right).
\end{equation}
More specifically, we consider the rate of purified Bell pair generation, defined as the number of high fidelity Bell pairs that can be distilled per second from the network. We compute this rate as
\begin{equation}\label{HashingRate}
R_H = Y_H \,\times \, \frac{dn}{dt},
\end{equation}
where $\frac{dn}{dt}$ is the rate at which imperfect Bell pairs are generated across the quantum network. It is defined as
\begin{equation}
    \frac{dn}{dt} = \frac{1}{T_{clock}} 
\end{equation}
where $T_{clock}$ represents the total time taken to distribute a Bell pair across the network. Together, Eqs.\ \eqref{Fidelity} and \eqref{HashingRate} determine the overall success and efficiency of entanglement distribution and distillation in an $n$-qubit repeater network. 

In Section~\ref{sec:effic-analysis}, we analyze both parallel and sequential entanglement distribution protocols. For parallel entanglement distribution, all intermediate nodes (excluding the receiver node) simultaneously and independently attempt to establish entanglement with their respective neighboring nodes. Once a node has been successfully entangled with its neighbors, it performs an entanglement swapping operation between the neighboring nodes. Conversely, a sequential entanglement distribution protocol builds entanglement iteratively beginning from the initial sender node. Each node attempts to generate entanglement with its neighboring node only after the previous link has been successfully established. Similar to the parallel case, a node performs entanglement swapping once it has become entangled with both of its neighbors.

In this section, we introduced success metrics for evaluating the performance of the entanglement distribution protocols for quantum repeater networks. We used fidelity as a closeness measure between the network output state and the desired Bell state, and hashing rate to quantify the efficiency of distilling Bell-type entanglement from imperfect Bell pairs produced by the quantum network. We reviewed the respective constructions of parallel and sequential entanglement distribution protocols, which we now simulate in the following section.

\section{Efficient Analysis of Asynchronous Entanglement Distribution}\label{sec:effic-analysis}

In this section we simulate entanglement distribution using the parallel and sequential protocols, incorporating the noise model derived in Section~\ref{NoiseModel}. Leveraging insight from our analytical result in Eq.\ \eqref{eq:fid-general}, we observe that the fidelity of the final Bell state depends on several constant parameters specific to a given network setup, namely the number of Bell state measurements $m$, the BSM ideality parameter $\lambda_{\text{BSM}}$, and the memory dephasing time $T_{\text{dp}}$. Crucially, among these factors, only the cumulative time that qubits spend in memory $T$, varies across different runs of the protocol in the same network. As a result, we can significantly simplify our analysis by computing just two key quantities: the total quantum memory time $T$ to calculate fidelity, and the overall protocol duration $T_{\text{clock}}$ to calculate hashing rate. We present and analyze the results of our analysis, and examine the role of imperfect Bell state measurement in each case. Finally, we compare the strengths and limitations of each protocol, and highlight their respective advantages and drawbacks for practical entanglement distribution in near-term quantum networks.

The analysis in this work are conducted under a base set of assumptions. We outline the key assumptions below that define our analytical framework.
\begin{enumerate}
    \item \textbf{Network Architecture:} Each repeater node is equipped with two quantum memories, one to store the photon generated by the node itself, and one to store the photon generated by its neighbor. Each end node has only one quantum memory: the sender stores only the photon it generates, and the receiver stores only the photon it receives from its neighbor. Every node, except for the receiver node, is equipped with an entangled photon source. 
    \item \textbf{Network Homogeneity:} The quantum network is homogeneous, consisting of evenly spaced repeater nodes with identical memory dephasing times.
    \item \textbf{No global knowledge:} There is no global knowledge of the network: each node operates autonomously and without an awareness of the full network structure or the state of distant nodes. Instead, each node operates using only the information it can access locally, i.e. the state of its own quantum memory and any classical communication received from its immediate neighbors.
    \item \textbf{Timing and Communication Assumptions:} Entangled photon generation and Bell state measurement, within a given node, are assumed to occur instantaneously and deterministically. Photons travel between nodes through optical fibers at $c = 2 \times 10^8$ m/s, and reach their neighboring nodes probabilistically. Classical communication is deterministic.
\end{enumerate}

We track the entanglement links between nodes in the quantum network state using a Boolean array of length $N-1$, where $N$ is the total number of nodes. Each array element corresponds to a link between adjacent nodes, and carries a value ``True'' if entanglement has been successfully established between the links adjacent nodes. For our analysis, time is normalized to $L/c$, where $L$ is the internode distance and $c$ is the speed of light in optical fiber. We first consider only the error due to dephasing in quantum memory. Since dephasing and BSM errors are independent this simplification is justified. In Section~\ref{sec:bsm-error}, we extend our analysis to include the impact of BSM imperfections.

Given that the final Bell state fidelity depends solely on the cumulative time qubits spend in memory, we can characterize the performance of both the sequential and parallel distribution protocols by analyzing the total quantum memory time $T$ and the overall protocol duration $T_{\text{clock}}$. The latter is required specifically for calculating the hashing rate. We now detail how both $T$ and $T_{\text{clock}}$ are computationally obtained in the analysis, and provide an algorithmic implementation for each protocol.  All code used in this work is publicly available online~\cite{hughes2025noisynet}.

\paragraph{Sequential Entanglement Distribution}\label{sec:sequential-wt}
Sequential distribution protocols can offer several advantages. For one, sequential protocols carry a lower resource requirement since fewer operations need to occur simultaneously. Additionally, global management of quantum memory is simplified since qubits are not typically left idle for any long periods of time. Moreover, errors arising throughout the distribution process can be easier to identify, and node timing more straightforward to synchronize. However, some drawbacks of sequential distribution protocols include longer overall entanglement distribution times, and an increased vulnerability to individual link failure. 

Algorithm~\ref{alg:sequential} below gives an programmatic instruction for computing parameters $T$ and $T_{\text{clock}}$ in the sequential protocol. We define the timing parameters used in the sequential protocol as follows:
\begin{itemize}
    \item $T_{\text{clock-step}}= 2$: Entanglement generation events take $2$ time steps to complete.
    \item $T_{\text{success}}= 3$: Cumulative memory time acquired by successfully generating an entanglement link. When generating entanglement between $A$ and $B$, Node $A$ adds $2$ time steps to memory, Node $B$ adds $1$ time step (Section~\ref{NoiseModel}).
    \item $T_{\text{existing}}= 4$: The memory time acquired by two existing qubits while an entanglement generation attempt is taking place.
\end{itemize}
\begin{algorithm}[ht]
\caption{Sequential Entanglement Distribution}
\label{alg:sequential}
\begin{algorithmic}[1]
\State $T = 0$
\State $T_{\text{clock}} = 0$
\State \texttt{links} $=$ [False] $\times$ $(N - 1)$
\For{\texttt{link} in \texttt{links}}
    \While{\texttt{link} is False}
        \If{\Call{RandomSuccess}{$p_{\text{loss}}$}}
            \State $T \mathrel{+}= T_{\text{success}}$
            \State \texttt{link} $=$ True
            \If{\texttt{link} is \textbf{not} first}      
                \State $T\mathrel{+}=T_{\text{existing}}$
            \EndIf
            
        \EndIf
    \State $T_{\text{clock}} \mathrel{+}= T_{\text{clock-step}}$
    \EndWhile
\EndFor
\State \Return $(T, \,T_{\text{clock}})$
\end{algorithmic}
\end{algorithm}

The sequential procedure iteratively attempts to establish entanglement between each pair of adjacent nodes along a linear chain of repeaters. At each step, entanglement is attempted until successful, the probability of successful entanglement generation is governed by a loss parameter $p_{\text{loss}}$. If successful, a memory time of $T_{\text{success}}$ is added to the cumulative memory time $T$. For all links except the first, an additional penalty $T_{\text{existing}}$ is added to account for decoherence of the two qubits held in memory as the entanglement attempt is taking place. The clock is incremented by $T_{\text{clock-step}}$ per attempt. The algorithm terminates once all $N-1$ links are successfully established, returning the total memory time $T$ and the total elapsed time $T_{\text{clock}}$.

\paragraph{Parallel Entanglement Distribution}\label{sec:parallel-wt}
In the parallel entanglement distribution protocol all nodes attempt to establish entanglement across multiple links simultaneously, regardless of the outcome from other links. The entanglement process occurs in \textit{rounds}, with each round consisting of entanglement attempts across all unestablished links. Successfully entangled links are indicated complete, and decoherence is determined by the number of qubits stored in memory, corresponding to the number of runs returning ``True'' values. Once all links are established the protocol terminates. Parallel entanglement protocols enable faster distribution on average, given the reduced overall entangling time. Accordingly, parallel distribution is especially well-suited for larger quantum networks with many nodes or longer distances between neighboring nodes. However, this ability to generate entangled pairs more rapidly comes at the cost of increased resource overhead, as multiple nodes and links must operate simultaneously. Coordinating entanglement swapping operations also becomes more complex, which can result in additional error and reduced overall fidelity. Moreover, the increased time qubits spend in memory leads to a greater risk of decoherence.

The main parallel protocol is detailed in Algorithm~\ref{alg:parallel}, and relies on the helper subroutines defined Algorithms~\ref{alg:parallel-round}, \ref{CountRuns}, and \ref{LinkSwap} to obtain $T$ and $T_{\text{clock}}$.

\begin{algorithm}[H]
\caption{Parallel Entanglement Distribution}
\label{alg:parallel}
\begin{algorithmic}[1]
\State $T = 0$
\State $T_{\text{clock}} = 0$
\State \texttt{links} $= [\text{False}] \times (N - 1)$
\While{not \Call{AllSwapped}{\texttt{links}}}
    \State $(\texttt{links}, T_{\text{round}}) = \Call{ParallelRound}{\texttt{links}, p_{\text{loss}}}$
    \State $T \mathrel{+}= T_{\text{round}}$
    \State $T_{\text{clock}} \mathrel{+}= T_{\text{clock-step}}$
\EndWhile
\State \Return $(T, \, T_{\text{clock}})$
\end{algorithmic}
\end{algorithm}
\begin{algorithm}[H]
\caption{Single Parallel Entanglement Round}
\label{alg:parallel-round}
\begin{algorithmic}[1]
\Function{ParallelRound}{\texttt{links}, $p_\text{loss}$}
    \State $T_\text{round} = 4 \; \times$ \Call{CountRunsOfTrue}{\texttt{links}}
    \For{\texttt{link} in \texttt{links}}
        \If{\texttt{link} = False}
            \If{\Call{RandomSuccess}{$p_\text{loss}$}}
                \State \texttt{link} = True
                \State $T_\text{round} \mathrel{+}= T_{\text{success}}$
            \EndIf
        \EndIf
    \EndFor
    \State \Return $(\texttt{links}, \,T_\text{round})$
\EndFunction
\end{algorithmic}
\end{algorithm}
\begin{algorithm}[H]
\caption{Count Runs of True}
\label{CountRuns}
\begin{algorithmic}[1]
\Function{CountRunsOfTrue}{\texttt{links}}
    \State \Return Number of contiguous runs of \texttt{True} in \texttt{links}
\EndFunction
\end{algorithmic}
\end{algorithm}
\begin{algorithm}[H]
\caption{All Links Swapped Check}
\label{LinkSwap}
\begin{algorithmic}[1]
\Function{AllSwapped}{\texttt{links}}
    \State \Return True if all elements in \texttt{links} are \texttt{True}
\EndFunction
\end{algorithmic}
\end{algorithm}
In the parallel scheme, all links between adjacent nodes in a linear chain attempt entanglement generation simultaneously during each round. The probability of successful entanglement generation is again governed by the loss parameter $p_{\text{loss}}$. In successful entanglement generation events, the link value is updated to ``True'', and a time of $T_{\text{success}}$ is added to the cumulative memory time $T$. An additional memory time of $4$ units is added to $T_\text{round}$ for each run of established links to account for two qubits held in memory per round of entanglement generation attempts. The total clock time $T_{\text{clock}}$ increases by a fixed increment $T_{\text{clock-step}}$ per round. The protocol repeats until all $N-1$ links have been successfully established, returning both the total memory time $T$ and the total elapsed time $T_{\text{clock}}$.

\subsection{Comparison of Sequential and Parallel Protocols}\label{ComparisonSection}

\begin{figure}[H]
    \centering
    \begin{subfigure}{0.371111111\linewidth}
        \centering
        \includegraphics[width=\linewidth]{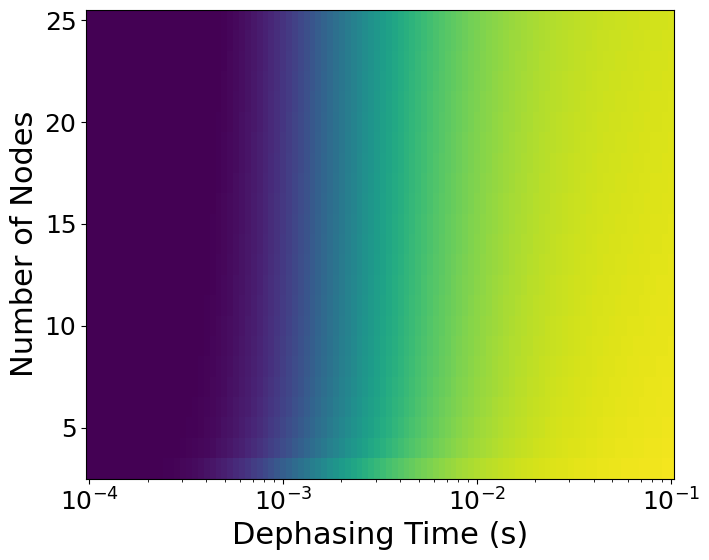}
        \caption{Sequential Fidelity}
        \label{fig:seq-fid}
    \end{subfigure}
    \begin{subfigure}{0.40\linewidth}
        \centering
        \includegraphics[width=\linewidth]{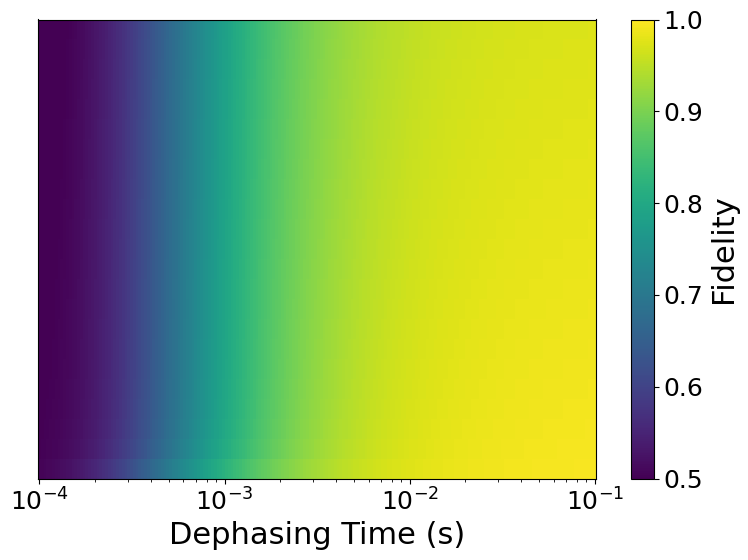}
        \caption{Parallel Fidelity}
        \label{fig:par-fid}
    \end{subfigure}
    
    \vspace{1em}
    
    \begin{subfigure}{0.371111111\linewidth}
        \centering
        \includegraphics[width=\linewidth]{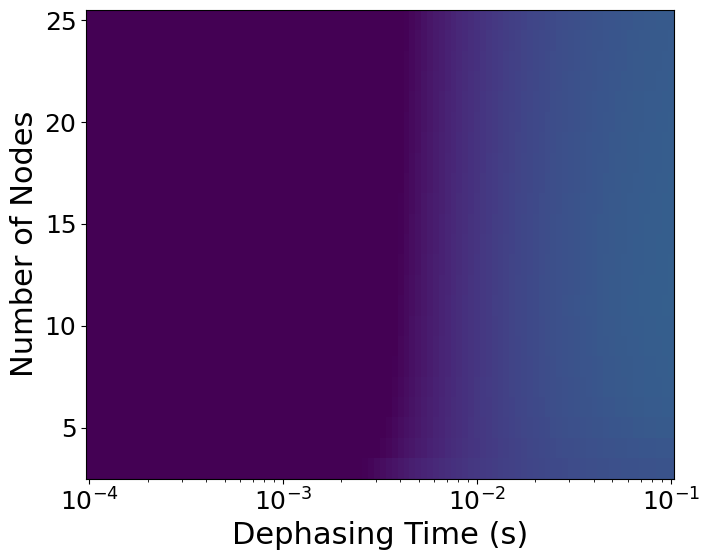}
        \caption{Sequential Hashing Rate}
        \label{fig:seq-hash}
    \end{subfigure}
    \begin{subfigure}{0.40\linewidth}
        \centering
        \includegraphics[width=\linewidth]{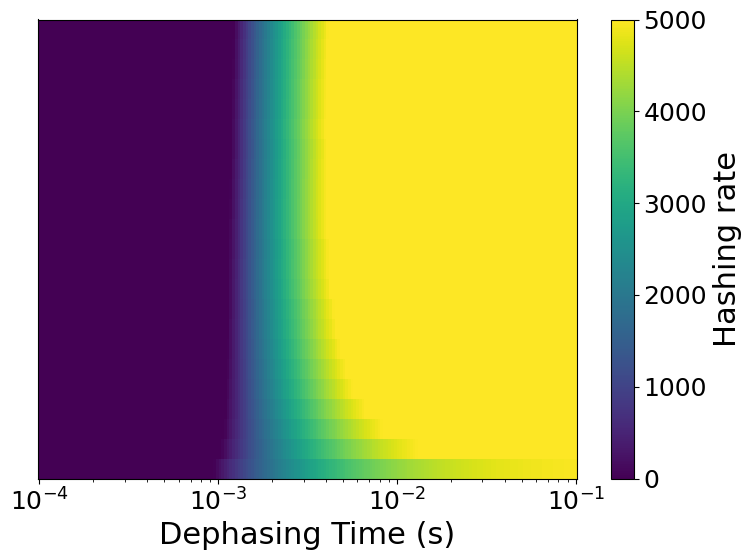}
        \caption{Parallel Hashing Rate}
        \label{fig:par-hash}
    \end{subfigure}
    \caption{Comparisons between asynchronous distribution protocols, conducted using fidelity and hashing rate metrics. Protocols are compared in networks with attenuation length $L_{att} =22$km and internode distance $L=50/N$km, with varying dephasing times $T_{dp}$ and number of nodes $N$. Subfigures (a) and (b) give the fidelity of distributed Bell pairs in the sequential and parallel protocols, respectively. Subfigures (c) and (d) depict corresponding hashing rates. While fidelity is comparable across both protocols, the parallel approach demonstrates a strong advantage in hashing rate.}
    \label{fig:protocol-comparison}
\end{figure}

We now evaluate and compare the respective performances of sequential and parallel asynchronous protocols in a linear quantum network of total length $50$ km and attenuation length $L_{att}=22$ km. Throughout our analysis, we vary both the number of nodes and the memory dephasing time. As discussed in Section~\ref{sec:metrics}, key metrics for success are the final fidelity of the distributed Bell state and the hashing rate. The results of our analysis are compiled in Figures~\ref{fig:protocol-comparison} and~\ref{fig:hashing-rates}.

We begin by comparing the output fidelity of the distributed Bell pair and the corresponding hashing rates for the sequential and parallel protocols. Figures~\ref{fig:seq-fid} and~\ref{fig:par-fid} illustrate how output fidelity varies with the number of nodes and the memory dephasing time for the sequential and parallel cases, respectively. Similarly, Figures~\ref{fig:seq-hash} and~\ref{fig:par-hash} give the associated hashing rates for each protocol.

When analyzing the output fidelity of distributed Bell pairs (Figures~\ref{fig:seq-fid} and~\ref{fig:par-fid}), both the sequential and parallel protocols exhibit similar performance, with the parallel protocol offering a slight advantage in fidelity. However, as illustrated in Figures~\ref{fig:seq-hash} and~\ref{fig:par-hash}, the parallel protocol achieves significantly higher hashing rate due to the reduced runtime and improved efficiency of entanglement generation and swapping events. While the sequential protocol typically holds fewer qubits in memory at any given time, at most two, its longer execution time results in cumulative decoherence comparable to that of the parallel protocol. Conversely, the parallel protocol, despite storing up to $N$ qubits simultaneously in memory, experiences a shorter overall execution time which serves to mitigate decoherence and improve overall performance.

In this section, we presented an analysis comparing the performance of sequential and parallel asynchronous entanglement distribution protocols. We evaluated the output fidelity of distributed Bell pairs, with increasing network size and increasing quantum memory dephasing time, as well as the hashing rates achieved by each protocol. While both approaches produced similar fidelities, the parallel protocol demonstrated a clear advantage in hashing rate due to its overall shorter runtime. In the next section we consider the effect of imperfect Bell state measurement on entanglement distribution, and assess its impact on fidelity and hashing rate.

\subsection{Impact of Bell State Measurement Imperfections}\label{sec:bsm-error}

We now investigate the effect of imperfect Bell state measurement on the performance of sequential and parallel distribution protocols. As shown in Eq.\ \eqref{eq:fid-general}, the final output fidelity scales exponentially with the number of BSMs, via the ideality parameter $\lambda_{\text{BSM}}$. We assess each protocol's tolerance to imperfect BSM by analyzing the hashing rate across a range of ideality parameters $\lambda_{\text{BSM}}$. As before, both protocols are evaluated over a range of network sizes and memory dephasing times.

\begin{figure}[ht]
    \centering
    \begin{subfigure}{0.315\linewidth}
        \centering
        \includegraphics[width=\linewidth]{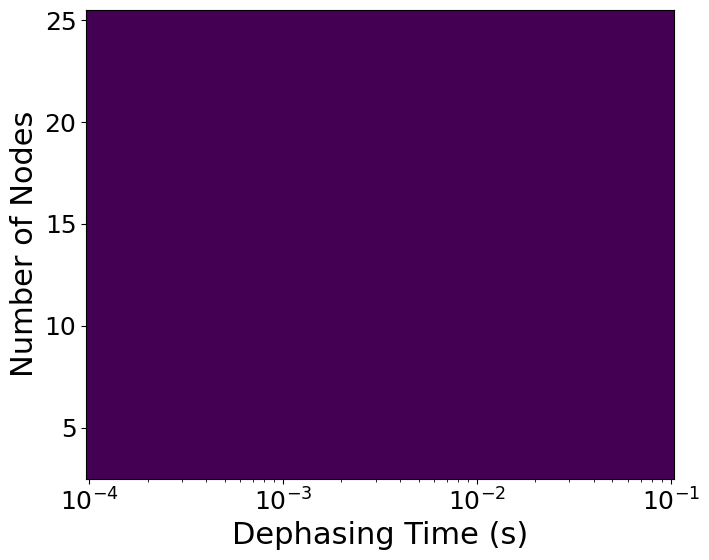}
        \caption{Seq., $\lambda_{BSM} = 0.9$}
        \label{fig:par-BSM9}
    \end{subfigure}
    \begin{subfigure}{0.293\linewidth}
        \centering
        \includegraphics[width=\linewidth]{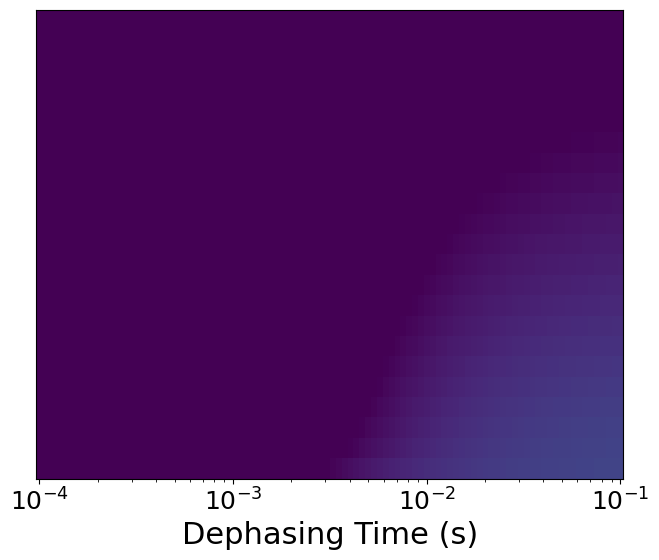}
        \caption{Seq., $\lambda_{BSM} = 0.99$}
        \label{fig:par-BSM99}
    \end{subfigure}
    \begin{subfigure}{0.34\linewidth}
        \centering
        \includegraphics[width=\linewidth]{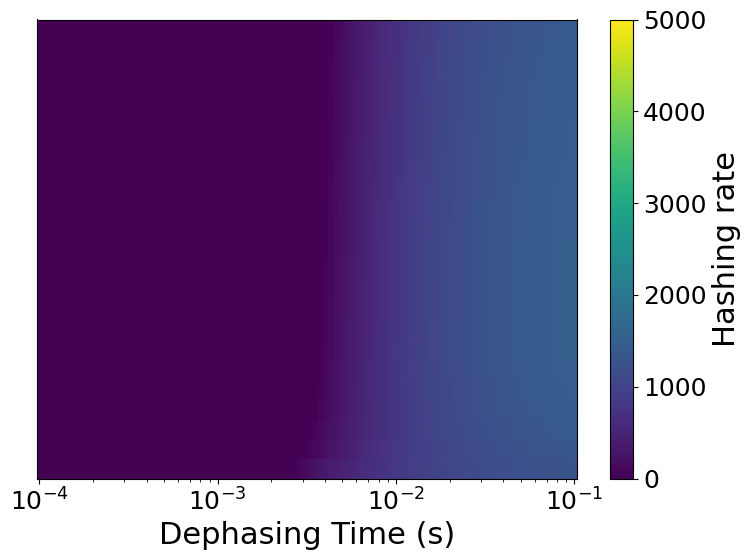}
        \caption{Seq., $\lambda_{BSM} = 0.999$}
        \label{fig:par-BSM999}
    \end{subfigure}
    \begin{subfigure}{0.315\linewidth}
        \centering
        \includegraphics[width=\linewidth]{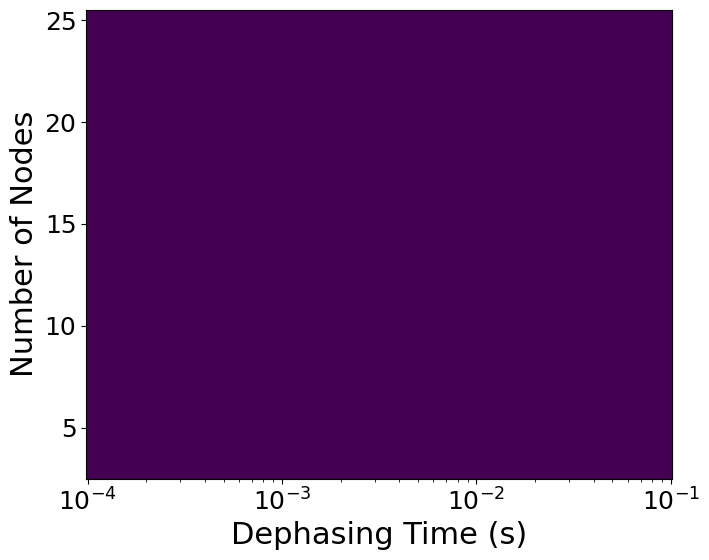}
        \caption{Par., $\lambda_{BSM} = 0.9$}
        \label{fig:seq-BSM9}
    \end{subfigure}
    \begin{subfigure}{0.293\linewidth}
        \centering
        \includegraphics[width=\linewidth]{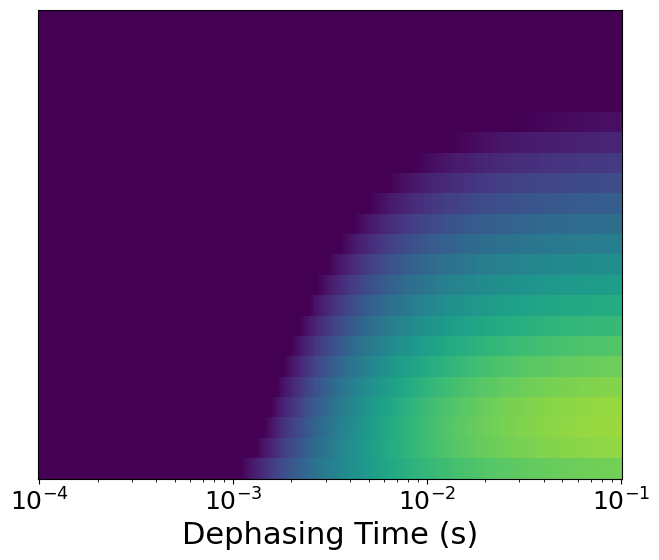}
        \caption{Par., $\lambda_{BSM} = 0.99$}
        \label{fig:seq-BSM99}
    \end{subfigure}
    \begin{subfigure}{0.34\linewidth}
        \centering
        \includegraphics[width=\linewidth]{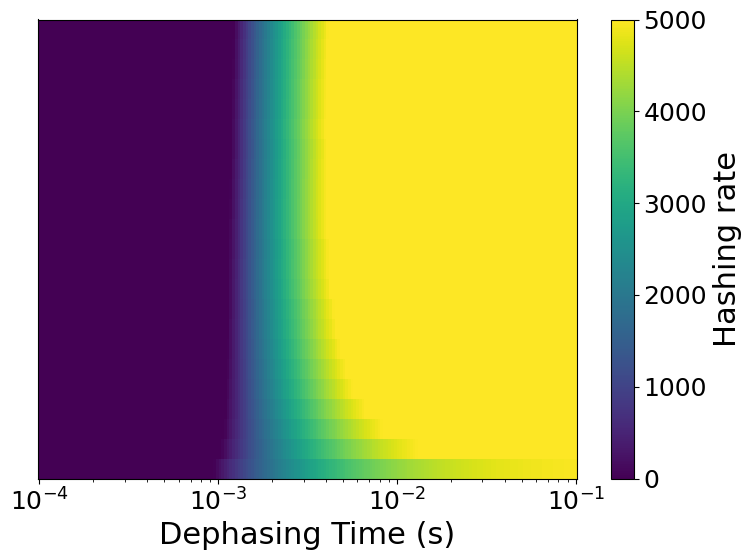}
        \caption{Par., $\lambda_{BSM} = 0.999$}
        \label{fig:seq-BSM999}
    \end{subfigure}

    \caption{Comparisons of imperfect BSM tolerance in sequential and parallel protocols. Protocols are compared in networks with attenuation length $L_{att} =22$km and internode distance $L=50/N$km, with varying dephasing times $T_{dp}$ and number of nodes $N$. Subfigures (a--c) give the hashing rate for the sequential protocol with increasing $\lambda_{\text{BSM}}$. Subfigures (d--f) show results for the parallel protocol. For sufficiently high $\lambda_{\text{BSM}}$, the parallel protocol is far more robust to imperfect BSM than the sequential.}
    \label{fig:hashing-rates}
\end{figure}

Figure~\ref{fig:hashing-rates} presents the hashing rate achieved by each protocol in the analysis as increasing values of $\lambda_{\text{BSM}}$. Figures~\ref{fig:seq-BSM9}--~\ref{fig:seq-BSM999} show the impact of increasing $\lambda_{\text{BSM}}$ on hashing rate for the sequential protocol, while Figures~\ref{fig:par-BSM9}--~\ref{fig:par-BSM999} display the corresponding results for the parallel protocol.

Figure~\ref{fig:hashing-rates} illustrates that the parallel entanglement distribution protocol once again significantly outperforms its sequential counterpart due to the faster runtime of the protocol as discussed in the previous subsection. However, when the ideality parameter is slightly reduced below a certain threshold both protocols fail to yield any purified Bell pairs, emphasizing the sensitivity of entanglement distribution to measurement errors in both protocols . As shown for the case when $\lambda_{\text{BSM}} = 0.9$, even minor imperfections in BSM fidelity can compound across longer chains, severely limiting overall performance. These results reiterate the importance of high-fidelity BSM for the development of near-term quantum networks.

In this section, we presented algorithms for simulating the asynchronous distribution of Bell pairs using both sequential and parallel protocols in noisy quantum networks, based on the noise model derived in Section~\ref{NoiseModel}. We evaluated the performance of each protocol, for networks up to $25$ nodes and quantum memory dephasing times ranging from $10^{-4}$ to $10^{-1}$ seconds. In each experiment we compared the output fidelity of the final distributed Bell pairs, as well as the hashing rate achieved by each protocol. We found that the sequential protocol, although requiring fewer qubits be stored in memory and thereby reducing individual quantum memory decoherence, necessitates a longer overall execution time, which results in alternative decoherence from the extended runtime. In contrast, the parallel protocol utilizes numerous quantum memories simultaneously, but benefits from a significantly faster execution. 

Despite differences in runtime-induced decoherence, both protocols produced comparable output Bell state fidelities. However, the parallel distribution protocol achieved significantly higher hashing rates, especially with the added effect of imperfect Bell state measurements, due to its shorter overall execution time. Notably, both approaches required a minimum BSM ideality threshold $\lambda_{\text{BSM}}$ to yield any purified Bell pairs at all. Distributed Bell state fidelities were found to be highly sensitive to even minor reductions in $\lambda_{\text{BSM}}$, emphasizing the importance of high-fidelity BSM for reliable entanglement distribution in large or noisy quantum networks. In the next section we conclude this paper and discuss future applications of this work.

\section{Discussion}

In this work, we present a compact analytical framework for evaluating contemporary asynchronous entanglement distribution protocols in linear quantum networks. By reducing the simulation of complex quantum processes to simple memory-time calculations, our approach captures the essential physics of noise and decoherence while remaining computationally efficient. We begin by identifying and modeling the dominant sources of error in near-term quantum hardware, which we use to derive a generic and scalable error description of Bell state distribution fidelity. One important result from our error analysis is the finding that the fidelity of a Bell state distributed across a quantum network depends only on the total quantum memory time experienced by all qubits, rather than how memory times are distributed among individual qubits. This insight underpins the simplicity and scalability of our framework for both sequential and parallel protocols, providing an accessible and transparent tool for analyzing entanglement distribution protocols efficiently under realistic noise conditions.

Using our analytical framework, we evaluate the performance of leading asynchronous entanglement distribution protocols: sequential distribution, where nodes establish entanglement one at a time along the repeater line, and parallel distribution, where all nodes attempt simultaneous and independent entanglement generation. We analyze the performance of each protocol by evaluating the fidelity of the distributed Bell pair, as well as the hashing rate (the rate at which purified Bell pairs can be produced). We analyze and compare each protocol across a range of network sizes and different quantum memory dephasing times, and determine that parallel distribution outperforms sequential distribution in all cases. While final Bell state fidelity was comparable in both protocols, the significantly shorter runtime of the parallel protocol resulted in much higher hashing rates. Moreover, when incorporating the effects of imperfect Bell state measurement we find that both protocols exhibit a strong sensitivity to reduced measurement fidelity. Our findings suggest that parallel asynchronous protocols are particularly well-suited for application in near-term quantum networks.

In this work, we focused on key sources of error in asynchronous entanglement distribution protocols, including optical loss, quantum memory decoherence, and imperfect Bell state measurement. In contrast to previous studies that compute fidelity via detailed memory timing histories~\cite{cisco-analysis-asynchronous}, our framework offers a lightweight analytical approach based solely on total memory time and the number of swapping operations. Expressing the output fidelity solely as a function of total memory time, we reveal the fundamental scaling behavior governing asynchronous protocols. This streamlined formulation enables efficient network analysis at significantly larger qubit scales.

Future extensions of this analytical framework could incorporate additional sources of error, such as detector dark counts and imperfect Bell state preparation. We anticipate these effects can be integrated into our existing noise model through multiplicative corrections to the final fidelity expression in Eq.\ \eqref{eq:fid-general}, thereby broadening the scope of errors that can be effectively captured through simple memory time calculations.

Both the sequential and parallel entanglement distribution protocols can benefit from additional refinements such as memory cutoff and entanglement purification to improve performance. A memory cutoff strategy automatically discards states that have remained in quantum memory beyond a specified threshold, thereby avoiding the accumulation of decoherence that would degrade output fidelity. Similarly, entanglement purification affords sender and receiver access to local operations and classical communication (LOCC), enabling higher quality Bell pairs to be distilled from a larger set of output states. Integrating these techniques would improve the fidelity of distributed states and allow for a more realistic and nuanced comparison of both protocols under practical constraints.

Our analytical framework can be directly adapted to study traffic management and routing challenges in multiparty quantum networks. One primary requirement for distributed quantum computation is the ability to distribute GHZ-type entanglement across spatially separated nodes~\cite{Hein:2006uvf,Negrin:2024tyj,Abane:2024ipy,hu2024dynamic,chang2023entanglement}. However, many existing entanglement routing protocols are developed in the absence of realistic noise or hardware constraints. As demonstrated in Appendix~\ref{app-N-qubits}, our framework supports the distribution of $n$-qubit GHZ states, enabling an analysis of GHZ-type entanglement routing to be conducted under practical noise conditions. By analyzing concurrent entanglement distribution requests across various network topologies, our framework can be used to design and benchmark scheduling algorithms, assess resource consumption, and identify performance bottlenecks. Each of these elements are essential for the successful implementation and scalability of quantum network architectures.

In this work we specifically focus on the noise affecting entanglement distribution in near-term quantum networks. However, our framework can naturally be extended to study the distribution of additional quantum resources across distant nodes a network. One particularly important resource for establishing quantum advantage is quantum magic~\cite{Bravyi:2004isx,Oliviero:2022euv,Oliviero:2022bqm}, which captures the degree of non-stabilizerness in a quantum system and correlates directly with the intractability of classically simulating that system. Moreover, recent work has demonstrated that magic can exist non-locally~\cite{Bao_2022,Cao:2024nrx}, and can be encoded in the entanglement structure of the system. While magic is essential to achieve computational speedups for quantum algorithms, its role in quantum networks remains largely unexplored. In pursuit of understanding non-local magic distribution, we seek to utilize our analytical framework to explore the feasibility and complexity of distributing magic across a quantum network.

The quantum networks studied throughout this paper can be modeled using graph states~\cite{Hein:2006uvf}, where vertices represent qubits (or more generally collections of interacting qubits) and edges correspond to entanglement. Numerous techniques have been developed to analyze the generation and dynamics of entanglement in graph states~\cite{fan2024optimized,cavalcanti2009open,fischer2021distributing,Couch:2019zni,Burchardt:2023odi}, leveraging the underlying algebraic structure to derive exact constraints on entanglement possibilities~\cite{Munizzi:2023ihc, Keeler:2022ajf, Keeler:2023xcx, Latour:2022gsf}. In quantum networks, similar constraints naturally emerge from fundamental and practical limitations in entanglement generation and distribution. By combining our realistic analytical model with an algebraic analysis of graph state entanglement we can more accurately probe the potential for noisy quantum networks to achieve desired entanglement arrangements and generation rates~\cite{Couch:2019zni,Keeler:2023shl}, offering diagnostics for quantum network capabilities.
 
\section*{Acknowledgments}

The authors thank Nicolas Dirnegger, Roberto Negrin, Richard Ross, Peter Shanahan, Prof. Chee Wei Wong for helpful discussions. We gratefully acknowledge funding from the National Science Foundation (NSF) under NSF Award 2107265 ``U.S.-Ireland R\&D Partnership: Collaborative Research: CNS Core: Medium: A unified framework for the emulation of classical and quantum physical layer networks".

\begin{appendix}
\addtocontents{toc}{\protect\setcounter{tocdepth}{1}}

\section{Appendix}
\subsection{Dephasing Error on a Four Qubit Network}\label{app-dephase-4}

We derive an expression for the fidelity of a $\ket{\phi^+}$ Bell state produced via entanglement swapping between two initial $\ket{\phi^+}$ Bell pairs. 

We begin by modeling the effect of dephasing on a two-qubit Bell state, $(\ket{\phi^+}\bra{\phi^+})_{AB}$, where qubits $A$ and $B$ remain in memory for times $t_A$ and $t_B$, respectively. The evolution is captured using the following Kraus map:

\begin{align}
K_{00} &= \sqrt{(1 - p_A) \; (1 - p_B)} \; (I \otimes I) \\
K_{01} &= \sqrt{(1 - p_A) \; p_B}       \;(I \otimes Z) \\
K_{10} &= \sqrt{p_A \; (1 - p_B)}       \;(Z \otimes I) \\
K_{11} &= \sqrt{p_A \; p_B}            \;(Z \otimes Z)
\end{align}

Here, $p_X$ is the dephasing probability for qubit $X \in \{A, B\}$, given by:

\begin{equation}
    p_{X} = \frac{1 - \exp^{-t_{X}/T_{dp}}}{2}
\end{equation}

Applying this channel to the initial Bell state yields:

\begin{align}
    \rho_{AB} &= \sum^1_{i,\,j=0} \; K_{ij} \; \ket{\phi^+}\bra{\phi^+} \; K_{ij}^\dagger \\
    &= (1-p_A)\,(1-p_B) \; \ket{\phi^+}\bra{\phi^+} + (1 - p_A) \; p_B \; \ket{\phi^-}\bra{\phi^-} \\
    &+ p_A\,(1-p_B) \; \ket{\phi^-}\bra{\phi^-} + p_A\,p_B \; \ket{\phi^+}\bra{\phi^+} \\
\end{align}

Regrouping, the final state becomes:
\begin{equation}
    \rho_{AB} = \alpha_{AB} \; \ket{\phi^+}\bra{\phi^+} + \beta_{AB} \;\ket{\phi^-}\bra{\phi^-}
\end{equation}\label{eqn:2bps-dp}
where $\alpha_{XY} = 1 - p_X - p_Y + 2\,p_X \, p_Y$ and $\beta_{XY} = p_X + p_Y - 2\,p_X \, p_Y$.

Now, consider two such Bell states $\rho_{AB}$ and $\rho_{CD}$, where each qubit has spent different times in memory. The joint state is:
\begin{align}\label{eq:rho-init-ABCD}
    \rho_{ABCD} &= \rho_{AB} \otimes \rho_{CD} \\
    &= \big(\alpha_{AB} \, \ket{\phi^+}\bra{\phi^+} + \beta_{AB} \, \ket{\phi^-}\bra{\phi^-} \big) \, \otimes \, \big(\alpha_{CD} \, \ket{\phi^+}\bra{\phi^+} + \beta_{CD} \, \ket{\phi^-}\bra{\phi^-} \big)
\end{align}

To model a Bell state measurement (BSM) on qubits $B$ and $C$, we apply the projection operator:

\begin{equation}
    P = \text{I} \otimes \ket{\phi^+} \bra{\phi^+} \otimes \text{I}
\end{equation}

This corresponds to the event where the BSM outcome is $\phi^+$. Of course, in practice the BSM measurement can yield any of the four Bell states; but with local operations and classical communication (LOCC), we can adjust the system to our desired state associated with the $\phi^+$ BSM outcome.

\begin{align}
    \rho_{AD} &= \text{Tr}_{BC}\,(P \, \rho_{ABCD} \, P ^\dagger) \\
    &= (\alpha_{AB} \, \alpha_{CD} + \beta_{AB}\, \beta_{CD}) \,\ket{\phi^+}\bra{\phi^+} \, + \, (\alpha_{AB} \, \beta_{CD} + \beta_{AB}\, \alpha_{CD}) \,\ket{\phi^-}\bra{\phi^-}
\end{align}\label{eq:rho-AD}

The fidelity of the resulting $\ket{\phi^+}$ Bell pair is thus:
\begin{equation}
    F = \alpha_{AB} \, \alpha_{CD} + \beta_{AB}\, \beta_{CD}
\end{equation}

\subsection{Generalization to N-Qubit Networks}\label{app-dephase-n}
We extend the above result to a six-qubit system by performing an additional entanglement swap on a new Bell pair $(E, F)$. A BSM on qubits $D$ and $E$ results in a final Bell pair between $A$ and $F$:
\begin{align}
    \rho_{ADEF} &= \rho_{AD} \otimes \rho_{EF} \\
    &= \big(\alpha_{AD} \, \ket{\phi^+}\bra{\phi^+} + \beta_{AD} \, \ket{\phi^-}\bra{\phi^-} \big) \, \otimes \, \big(\alpha_{EF} \, \ket{\phi^+}\bra{\phi^+} + \beta_{EF} \, \ket{\phi^-}\bra{\phi^-} \big)
\end{align}

where 
$$
\alpha_{AD} = \alpha_{AB} \,\alpha_{CD} + \beta_{AB} \,\beta_{CD}
$$
$$
\beta_{AD} = \alpha_{AB} \,\beta_{CD} + \beta_{AB} \,\alpha_{CD}
$$

Performing a BSM on $(D, E)$, the resulting state of $(A, F)$ has fidelity:
\begin{equation}
    \rho_{ADEF} = \big(\alpha_{AD} \, \alpha_{EF} + \beta_{AD} \, \beta_{EF} \big) \, \ket{\phi^+}\bra{\phi^+} \,+\, \big(\alpha_{AD} \, \beta_{EF} + \beta_{AD} \, \alpha_{EF} \big) \, \ket{\phi^-}\bra{\phi^-}
\end{equation}

Substituting gives:
\begin{equation}
    F = \big(\alpha_{AB} \, \alpha_{CD} + \beta_{AB} \, \beta_{CD} \big) \, \big(\alpha_{EF} + \beta_{EF}\big) 
\end{equation}

\subsection{Imperfect Bell State Measurement}\label{app-bsm-error}

An imperfect Bell state measurement is modeled as a two-qubit depolarizing channel followed by an ideal BSM. For four qubits $(A, B, C, D)$, the joint state is:

\begin{equation}
    \rho_{ABCD} = (\alpha_{AB} \, \ket{\phi^+}\bra{\phi^+} + \beta_{AB} \,\ket{\phi^-}\bra{\phi^-}) \,\otimes\, (\alpha_{CD} \, \ket{\phi^+}\bra{\phi^+} + \beta_{CD} \,\ket{\phi^-}\bra{\phi^-})
\end{equation}

Applying the depolarizing channel with strength $\lambda_{BSM}$ and projecting onto the Bell basis:
\begin{equation}
    \begin{aligned}
        \rho &= \lambda_{BSM} \,\Big[
        (\alpha_{AB}\,\alpha_{CD} + \beta_{AB} \, \beta_{CD})\ket{\phi^+}\bra{\phi^+} 
        + (\alpha_{AB}\,\beta_{CD} + \beta_{AB} \, \alpha_{CD})\ket{\phi^-}\bra{\phi^-} \Big] \\
        &\quad + (1-\lambda_{BSM}) \,\Big[ \alpha_{AB}\,\alpha_{CD}\ket{++}\bra{++} + \alpha_{AB}\,\beta_{CD}\ket{+-}\bra{+-} \\
        &\hspace{3cm} + \beta_{AB}\,\alpha_{CD}\ket{-+}\bra{-+} + \beta_{AB}\,\beta_{CD}\ket{--}\bra{--} \Big] 
    \end{aligned}
\end{equation}

We have a mixture of two states. The first, with coefficient $\lambda_{BSM}$, is the state resulting from a perfect Bell state measurement with dephasing in memories obtained in Equation~\ref{eq:rho-AD}. The second, with coefficient $(1-\lambda_{BSM})$, is a fully mixed, classical state.

To expand this to a 6 qubit system, we consider adding an additional Bell pair to the system (denoted by qubits E, F) and performing an imperfect Bell state measurement with coefficient $\lambda_{BSM}'$ to qubits D, E, to establish a Bell pair between A, F. 

We split the density matrix into components such that
\begin{equation}
    \rho = \lambda_{BSM}\,\rho_1 + (1-\lambda_{BSM}) \, \rho_2
\end{equation}
where $\rho_1$ the entangled component and $\rho_2$ is the classical component. 

We will compute the outcome of Bell state measurements on the constituent parts of the density matrix separately. Firstly we will look at the entangled component $\rho_1$:

\begin{equation}
    \rho_1 \otimes \rho_{EF} = (\alpha_{AD} \, \ket{\phi^+}\bra{\phi^+} 
    + \beta_{AD} \, \ket{\phi^-}\bra{\phi^-})  \otimes  (\alpha_{EF} \, \ket{\phi^+}\bra{\phi^+} + \beta_{EF} \, \ket{\phi^-}\bra{\phi^-})
\end{equation}

This has the same form as the two Bell pairs in memory with no depolarizing error so we know after a Bell state measurement it will take the form:

\begin{equation}
    \begin{aligned}
        \rho &= \lambda_{BSM}' \,\Big[
        (\alpha_{AD}\,\alpha_{EF} + \beta_{AD} \, \beta_{EF})\ket{\phi^+}\bra{\phi^+} 
        + (\alpha_{AD}\,\beta_{EF} + \beta_{AD} \, \alpha_{EF})\ket{\phi^-}\bra{\phi^-} \Big] \\
        &\quad +  \, (1-\lambda_{BSM}') \,\Big[ \alpha_{AD}\,\alpha_{EF}\ket{++}\bra{++} + \alpha_{AD}\,\beta_{EF}\ket{+-}\bra{+-} \\
        &\hspace{3cm} + \beta_{AD}\,\alpha_{EF}\ket{-+}\bra{-+} + \beta_{AD}\,\beta_{EF}\ket{--}\bra{--} \Big]
    \end{aligned}
\end{equation}

Now looking at the classical component, $\rho_{2}$, we have:
\begin{equation}
\begin{aligned}
    \rho_2 \otimes \rho_{EF} &=\,\Big[ \alpha_{AB}\,\alpha_{CD}\ket{++}\bra{++} + \alpha_{AB}\,\beta_{CD}\ket{+-}\bra{+-} \\
    &+ \beta_{AB}\,\alpha_{CD}\ket{-+}\bra{-+} + \beta_{AB}\,\beta_{CD}\ket{--}\bra{--} \Big] \\ &\otimes (\alpha_{EF} \, \ket{\phi^+}\bra{\phi^+} + \beta_{EF} \, \ket{\phi^-}\bra{\phi^-})
\end{aligned}
\end{equation}

A Bell state measurement on this state, of course transfers no entanglement and we are left with the state $\rho_2$ across qubits A and F.

In total the state across A and F is:
\begin{equation}
    \begin{aligned}
        \rho &= \lambda_{BSM} \,\lambda_{BSM}' \,\Big[
        (\alpha_{AD}\,\alpha_{EF} + \beta_{AD} \, \beta_{EF})\ket{\phi^+}\bra{\phi^+} 
        + (\alpha_{AD}\,\beta_{EF} + \beta_{AD} \, \alpha_{EF})\ket{\phi^-}\bra{\phi^-} \Big] \\
        &\quad + \lambda_{BSM} \, (1-\lambda_{BSM}') \,\Big[ \alpha_{AD}\,\alpha_{EF}\ket{++}\bra{++} + \alpha_{AD}\,\beta_{EF}\ket{+-}\bra{+-} \\
        &\hspace{3cm} + \beta_{AD}\,\alpha_{EF}\ket{-+}\bra{-+} + \beta_{AD}\,\beta_{EF}\ket{--}\bra{--} \Big]\\
        &\quad + (1-\lambda_{BSM})  \,\Big[ \alpha_{AB}\,\alpha_{CD}\ket{++}\bra{++} + \alpha_{AB}\,\beta_{CD}\ket{+-}\bra{+-} \\
        &\hspace{3cm} + \beta_{AB}\,\alpha_{CD}\ket{-+}\bra{-+} + \beta_{AB}\,\beta_{CD}\ket{--}\bra{--} \Big]
    \end{aligned}
\end{equation}

The composition of the classical state does not matter for our purposes. We are only interested in the amplitude of the entangled state. We can see that each time an entanglement swapping operation with efficiency $\lambda_{BSM}$ is performed, the fidelity of the $\phi^+$ Bell state gains a factor of $\lambda_{BSM}$. Additionally, the errors of dephasing and depolarization are independent from each other; the magnitude of one has no impact on the magnitude of the other. Therefore, for a Bell pair generated from a network involving n entanglement swapping operations will have fidelity:
\begin{equation}
    F = \frac{(\lambda_{BSM})^n}{2}(1+e^{\frac{-T}{T_{dp}}})
\end{equation}

\subsection{Distributing an N-Qubit GHZ State}\label{app-N-qubits}

We now illustrate how entanglement swapping can distribute an $N$-qubit GHZ state. Consider an initial GHZ state shared between $N-1$ qubits labeled $A$, and one qubit $B$ at the sender:
\[
\ket{\Psi} = \frac{1}{\sqrt{2}} \left( \ket{0}^{\otimes N-1}_A \ket{0}_B + \ket{1}^{\otimes N-1}_A \ket{1}_B \right) \otimes \frac{1}{\sqrt{2}}(\ket{00}_{CD} + \ket{11}_{CD})
\]

To swap qubit $B$ into a distant node, a BSM is performed on qubits $B$ and $C$. Applying the projection operator:
\[
P = I \otimes \ket{\phi^+}\bra{\phi^+} \otimes I
\]

and tracing out $B$ and $C$, the final state of qubits $A$ and $D$ becomes:
\[
\ket{\Psi'} = \frac{1}{\sqrt{2}} \left( \ket{0}^{\otimes N-1}_A \ket{0}_D + \ket{1}^{\otimes N-1}_A \ket{1}_D \right)
\]

This process can be repeated to distribute the full GHZ state to multiple distant parties.

\end{appendix}

\bibliographystyle{JHEP}
\bibliography{noisynetworks}

\end{document}